\documentclass[]{elsart}
\usepackage{amsmath}
\usepackage{amssymb}
\usepackage{bbm}
\usepackage{xcolor}
\usepackage{booktabs}
\usepackage{algpseudocode}
\usepackage{algorithm}
\usepackage{theorem}
\usepackage[]{natbib}
\usepackage{tikz}
\usetikzlibrary{arrows}
\usepackage{enumitem}

\newcounter{count}

\newenvironment{assumption}[1][Assumption \arabic{count}]{\vspace{1em}\refstepcounter{count}\begin{trivlist}
\item[\hskip \labelsep {\bfseries #1}]}{\end{trivlist}\vspace{1em}}

\begin{document}

\begin{frontmatter}

\title{Mean-field theory of the general-spin Ising model}
\author{Lourens Waldorp, }
\author{Tuan Pham, and }
\author{Han L. J.  van der Maas}
\runauthor{Waldorp et al.}
\runtitle{Symmetric multivalued Ising modwel}
\address{University of Amsterdam, Nieuwe Achtergracht 129-B, 1018 NP, the Netherlands\\
{\tt waldorp@uva.nl}}

\begin{abstract}
Motivated by modelling in physics and other disciplines, such as sociology and psychology, we derive the mean field of the general-spin Ising model from the variational principle of the Gibbs free energy. The general-spin Ising model has $2k+1$ spin values, generated by $-(k-j)/k$, with $j=0,1,2\ldots,2k$, such that for $k=1$ we obtain $-1,0,1$, for example; the Hamiltonian is identical to that of the standard Ising model. The general-spin Ising model exhibits spontaneous magnetisation, similar to the standard Ising model, but with the location translated by a factor depending on the number of categories $2k+1$. We also show how the accuracy of the mean field depends on both the number of nodes and node degree, and that the hysteresis effect decreases and saturates with the number of categories $2k+1$. Monte Carlo simulations confirm the theoretical results. 

\end{abstract}

\end{frontmatter}
\endNoHyper

\section{Introduction}\noindent
The Ising model is a popular model, used in physics but also in computational science, econophysics, sociophysics and psychosociophysics \citep{Lee:2025,Castellano:2009,Jusup:2022,Maas:2020, Macy:2024}. 
The model is relatively simple, yet exhibits intriguing 
phase transitions, both first- and second order. To obtain such qualitative descriptions, we explore mean-field theory of a generalisation of the Ising model that extends beyond the spin values $-1$ and $1$ to include, for any $k\in \mathbb{N}$, the spin values generated by $(-k+j)/k$, with $j=0,1,2,\ldots,2k$. For example, with $k=1$, we get $\{-1,0,1\}$. This model is called the general-spin Ising model and has been introduced before in \citet{Rabe:1994}, where the interest was mostly in estimating parameters, and even earlier in \citet{Suzuki:1965}, which defined an Ising model for multiple states in general and derived the correlation functions. 

On the face of it, the general-spin Ising model seems related to the continuous XY model and a discretisation of the Heisenberg model. However, the general-spin Ising model shows very different behaviour than both the standard Heisenberg and XY model; the general-spin Ising model has a first-order phase transition while the standard Heisenberg and XY model do not, even in the mean-field limit  \citep{Barma:2022, Kirkpatrick:2013}.
The Blume-Capel model, which includes the value 0 and has an additional term for the energy of non-zeros (crystal field), has been investigated as a general-spin model \citep{Plascak:1993,Costabile:2014,Salama:2024}. The interest there was mostly in the effect of the crystal field term on magnetisation.  
Variations of the Ising model, such as the Potts model, are similar to the general-spin Ising model in that they involve multiple states; however, they only contribute to the energy when neighboring states are equal. \citep{Durrett:2007}; here any product of fractional values can contribute to the energy. The general-spin Ising model is different from the above models because it can be defined on the integers $\mathbb{Z}$, and here we normalise the values by the maximum $k$, so that all values are between -1 and 1. 

Applications of the general-spin Ising model in physics mostly involve systems with quantized spin values \citep{Costabile:2014} and mixed spin cases to model different magnetic materials \citep{Albayrak:2006,Deviren:2009,Espriella:2018}. For example, in \cite{Ertacs:2018} a two-layer square  lattice is used, where nodes within each layer represent one of the materials in thin film, and interactions within and between layers are modeled according to the Ising model with spin values $\pm 2$, $\pm 1$, $0$. However, our inspiration comes from the use of statistical physics models in sociology and psychology \citep{Maas:2024}. In sociophysics, for example, agents' opinions on social networks (e.g., Facebook, Instagram) may span a range of ordered values, such as from extreme left to extreme right, including a neutral position \citep[e.g.,][]{Chen:2015b}. Similarly, in mathematical psychology, the Ising model is applied to phenomena such as attitudes \citep{Dalege:2018} or major depression disorder \citep{Cramer:2016}, where node values can also vary along a spectrum, from negative to positive, again, including a neutral position. In both last two contexts, nodes are often measured by questionnaires, the most common of which are Likert scales. A typical Likert scale item asks respondents to rate their position on a 7-point scale ranging from 'strongly disagree' to 'strongly agree' \citep{Jebb:2021}. Currently, visual analogue scale responses (sliders) are becoming popular \citep{Betella:2016,Haslbeck:2025b} which can also be covered by the general-spin Ising model with large $k$. 

The Ising model is relevant to social sciences because it captures the interactions by the product of the variables. Known as the law of mass action \citep{Erdi:1989}, it can be found in research on human interactions, such as in opinion dynamics, e.g., the voter model \citep{Rredner:2019, Sen:2014} and in epidemiology, e.g., in susceptible-infected-susceptible models \citep{Keeling:2005}, but also in research on intelligence \citep{Savi:2019}, attitudes \citep{Dalege:2017}, and psychopathology \citep{Cramer:2010,Borkulo:2015,Waldorp:2019}.

Our interest is qualitative descriptions, and mean-field theory is appropriate for that purpose. Mean-field theory is correct in the thermodynamic limit (infinite graph size), but is still quite accurate when the mean degree of nearest neighbours is sufficiently high \citep{Gleeson:2012}. In physics this is often true, since the typical graph topology in physics is the $d$-dimensional lattice $\mathbb{Z}^d$ \citep[e.g., ][]{Aizenman:1987,Georgii:2001, Grimmett:2010}. Also in more modern applications of (variations of) the Ising model, versions of the $d$-dimensional lattice are used \citep{Ertacs:2018}. In social science applications, graph topologies often resemble a small world, and in neuroscience the topology often resembles a scale-free graph \citep{Sen:2014,Bassett:2006}, although sometimes the degree distribution resembles a Poisson distribution \citep{Chen:2015b}, suggesting an Erd\"{o}s-Reny\'{i} random graph. In applications to psychopathology, the graph topology appears to be sparse and cannot be distinguished from an Erd\"{o}s-Reny\'{i} random graph \citep{Borkulo:2015,Castro:2024}. 

We derive the mean-field equations for the general-spin Ising model with values between $-1$ and $+1$. 
After introducing the model and some notation in Section \ref{sec:multising}, we consider the mean field equations in Section \ref{sec:mean-field}. Then, in Section \ref{sec:phase-transitions} we show that spontaneous magnetisation depends on an additional factor containing the number of categories $2k+1$. In Section \ref{sec:numerical-illustration} we show with Monte Carlo simulations that the results of the Metropolis algorithm for the general-spin Ising model correspond well to the theoretical predictions,  including dependence on system size and the degree of nodes.

\section{General-spin Ising model}\label{sec:multising}\noindent
We have graph $\mathcal{G}$ consisting of $n$ nodes (vertices) in the set $V=\{1,2,\ldots,n\}$, and edges (connections) $E=\{(s,t):s,t\in V \, ;\, s \,{\rm and} \, t\, {\rm are\, neighbors}\}$. The topology we use is a random (Erd\"{o}s-Reny\'{i}) graph with probability of connecting $p_e$, independently and identically for each pair of nodes; this leads to on average $d=p_e(n-1)$ connections.  

\begin{figure}[ht]\centering
\pgfimage[width=.6\textwidth]{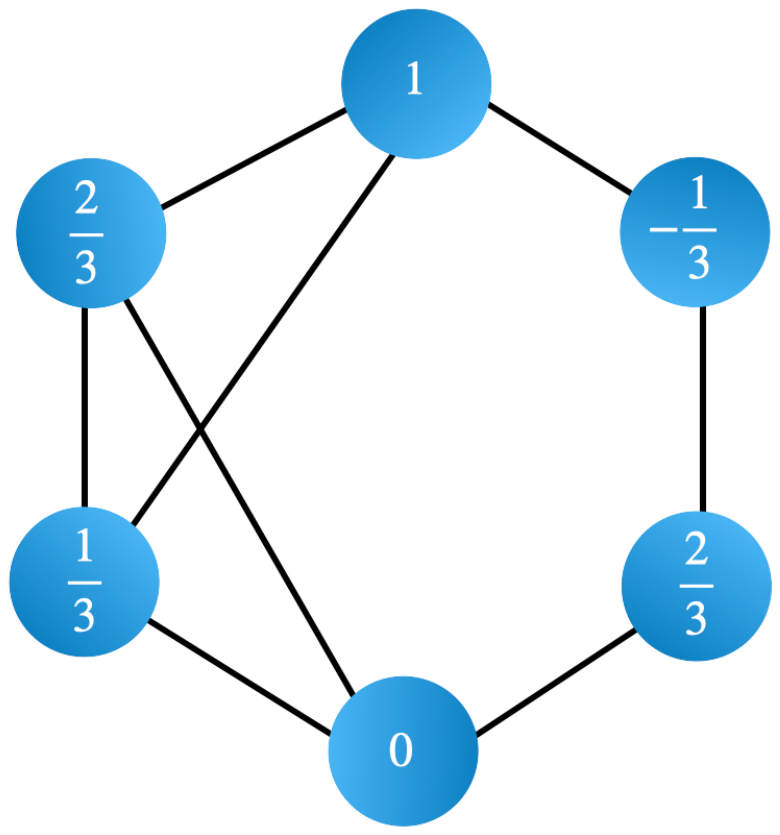}\vspace{-6em}
\caption{ General-spin Ising model with $k=3$, and so there are $7$ possible states.}
\label{fig:general-spin-pic}
\end{figure}

We assign to any node $s\in V$ a random variable $X_s$. The random variable $X_s$ can take values $x_s\in \Omega_k$ where $\Omega_k$ is generated by $(-k+j)/k$ for $j=0,1,2,\ldots,2k$, that is,
\begin{align*}
    \Omega_k:=\left\{\frac{-k}{k},\frac{-k+1}{k},\frac{-k+2}{k},\ldots,\frac{k-1}{k},\frac{k}{k}\right\},
\end{align*}
for any finite natural number $k$. Note that the regular Ising model can be obtained by taking $k=\tfrac{1}{2}$, i.e., $\Omega_{1/2}=\{-1,1\}$. Figure \ref{fig:general-spin-pic} shows an example with $k=3$ and $\Omega_3=\{-1,-\tfrac{2}{3},-\tfrac{1}{3},0,\tfrac{1}{3},\tfrac{2}{3},1\}$.

Given external field parameter $\tau\in \mathbb{R}$ and (ferromagnetic) interaction parameter $\sigma\ge 0$, the Hamiltonian is
\begin{align}
    \mathcal{H}(x) = -\tau \sum_{s\in V} x_s - \sigma \sum_{(s,t)\in E} x_s x_t,
\end{align}
where the sum over $(s,t)\in E$ runs over all edges. Throughout the paper, we shall refer to  the general-spin Ising model with this Hamiltonian and  $k\in \mathbb{N}$ as Ising$(k)$. At a temperature $T$, the probability of finding any realisation $x=(x_1,x_2,\ldots,x_n)$ of $X=(X_1,X_2,\ldots,X_n)$ is a Boltzmann probability and is
\begin{align}\label{eq:ising-model}
    \mathbb{P}_\theta(x) = \frac{1}{Z_\theta}\exp(-\beta\mathcal{H}(x)),
\end{align}
where $\beta=1/T$ is the inverse temperature, $\theta =(\beta, \sigma,\tau)$, and $Z_\theta$ is the partition function (normalising constant)  $Z_\theta:=\sum_{\{x\}} \exp(-\beta\mathcal{H}(x))$, where $\sum_{\{x\}}$ denotes the summation over all $(2k+1)^n$ possible configurations.

\section{Mean field}\label{sec:mean-field}\noindent
The mean field can be obtained by minimising the Gibbs free energy, which reveals the moments of the distribution \citep[see e.g.,][Chapter 3]{McCulloch:1988,Wainwright:2008}. Assuming weak correlations between variables, we obtain the mean-field Hamiltonian (see Appendix \ref{app:mean-field})
\begin{align*}
\mathcal{H}_{\mu}(x)=-\sigma\mu^2 n \frac{d}{2} -(\sigma\mu d +\tau)\sum_{s\in V} x_s,
\end{align*}
where $\mu$ is the mean field, the average effect from neighbouring nodes, and $d$ is the average degree of any node in the random graph $\mathcal{G}$. Using this mean-field Hamiltonian, the Gibbs free energy is 
\begin{align}\label{eq:free-energy}
G_\mu=-\frac{1}{\beta}\log Z_\mu =
\sigma\mu^2 n \frac{d}{2} - \frac{n}{\beta}\log\sum_{x\in \Omega_k^+}2\cosh\left(\beta(\sigma\mu d+\tau)x\right),
\end{align}
where $\Omega_k^+=\{0,\tfrac{1}{k},\ldots,1\}$.
Minimising $G_\mu$ we obtain the mean field $\mu$ (Appendix \ref{app:mean-field})
\begin{align}\label{eq:mean-field}
    \mathbb{E}(x) =\mu 
    &=\frac{\sum_{x\in \Omega_k^{+}}x\sinh(x\beta\gamma)}
    {\sum_{x\in \Omega_k^{+}}\cosh(x\beta\gamma)},
\end{align}
where $\gamma = \tau + \sigma \mu d$. The expression  of $\mu$ for Ising$(k)$ is similar in spirit to the one of the Ising model, except that we have a sum of terms ranging over $\Omega_k^+$. The susceptibility (variance) is obtained by taking the derivative of the mean field in (\ref{eq:mean-field}) with respect to $\tau$ \citep{Plischke:1994}. We obtain 
\begin{align}
    \mathbb{E}(x^2)-(\mathbb{E}(x))^2=\chi = \beta\left(\frac{\sum_{x\in \Omega_k^{+}}x^2\cosh(x\beta\gamma)}{\sum_{x\in \Omega_k^{+}}\cosh(x\beta\gamma)}-\mu^2\right).
\end{align}

\begin{figure}[ht]\centering
\begin{tabular}{@{\hspace{-1em}} c @{\hspace{1em}} c @{\hspace{-1em}} c}
		&\pgfimage[width=.55\textwidth]{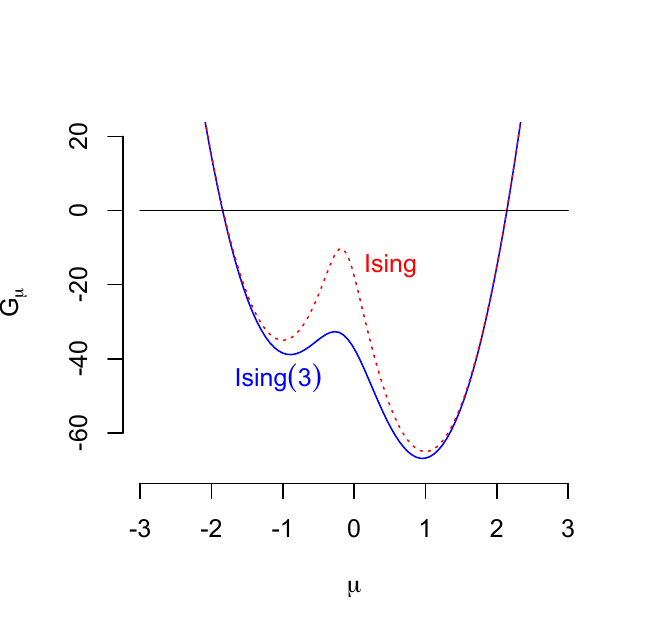}		&\pgfimage[width=.55\textwidth]{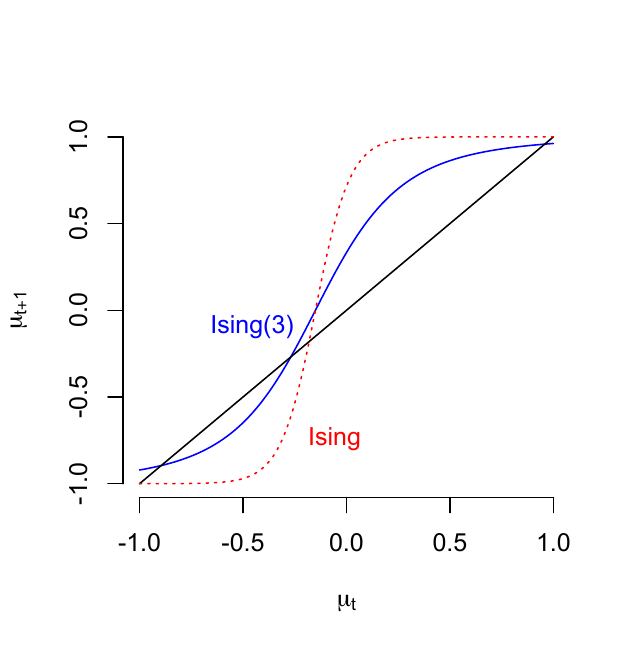}\\[-1em]
	&	(a)		&(b)
\end{tabular}
\caption{In (a) is the free energy function Eq. \eqref{eq:free-energy} with settings $\tau=0.3$, $d=2$, $\sigma=1$, $\beta=3$. For the general-spin Ising model $k$ is $3$ (blue) and the regular Ising model is shown for comparison (dashed, red). In (b) is an itereated map of the same general-spin Ising$(3)$ model as in (a) with $\beta=3$, and the regular Ising model with the same settings.  }
\label{fig:free-energy}
\end{figure}

Using the mean field we can investigate qualitatively some properties of the general-spin Ising model. 
Figure \ref{fig:free-energy}(a) shows the free energy of the mean field approach for the general-spin Ising model as well as for the regular Ising model. In this example, the general-spin Ising model has 7 spin values $-1,-\tfrac{2}{3},-\tfrac{1}{3},0,\tfrac{1}{3},\tfrac{2}{3},1$; we denote this version of the general-spin Ising model Ising$(3)$, since $k=3$. From Figure \ref{fig:free-energy}(a) three results are immediately clear. First, the number of local and global minima is the same for the Ising$(3)$ as for the regular Ising model. Second, the perturbation necessary to switch to a lower minimum is smaller for the Ising$(3)$ model than for the regular Ising model as the barrier's height is lower. And third, the minima of the free energy of the general-spin Ising model appear closer to 0 than the minima of the regular Ising model. 

In Figure \ref{fig:free-energy}(b) we see an iteration map that can be used for a graphical analysis of the fixed points of the general-spin Ising model. On the $x$-axis we have the value of the mean field in Eq. \eqref{eq:mean-field} at time $t$ and on the $y$-axis we see the value of the mean field at time $t+1$; we see a single iterate of the mean-field function. Both the general-spin Ising model with $k=3$ and the regular Ising model show three fixed points, where the middle fixed point is repelling and the two extreme fixed points are attracting. It also shows that the fixed points, corresponding to the minima of the free energy, are a bit closer to 0 for the general-spin Ising model than for the regular Ising model. 
%

\section{Phase transitions}\label{sec:phase-transitions}\noindent
We first discuss the phase transition of the magnetisation $\mu$ as a funciton of the inverse temperature $\beta$,  without an external field. By using Landau theory \citep[e.g.,][]{Plischke:1994}, we obtain that the phase transition of the general-spin Ising model is similar to the (second-order) phase transition of the regular Ising model, but for the general-spin Ising model the location of spontaneous magnetisation is translated by a factor determined by the number of categories $2k+1$. In particular, the phase transition near spontaneous magnetisation $\mu=0$ is of order (see Appendix \ref{app:landau-theory})
\begin{align}\label{eq:phase-transition-temp}
    \mu\approx
    \begin{cases}
        \pm \sqrt{r\beta \sigma d-1}   
        &\text{ if } 
        r\beta \sigma d \searrow 1\\
        0 &\text{ if } 
        r\beta \sigma d \le 1
    \end{cases},
\end{align}
where $r=(2k+1)/6k$ and $d$ is the average degree of any node in the random graph $\mathcal{G}$. In terms of the critical temperature $\beta_c=1/r\sigma d$ we obtain $\mu \propto (\tfrac{\beta}{\beta_c}-1)^{1/2}$ if $\beta \searrow \beta_c$, and 0 otherwise. Interestingly, the factor $r$ is between $\tfrac{1}{2}$ when $k=1$ (with categories $-1$, $0$, and $1$), and $r=\tfrac{1}{3}$ for $k\to \infty$. Near criticality, the susceptibility, is approximately $\chi\approx \beta (r-\mu^2)$, for small $\mu$.

\begin{figure}[t]\centering
\begin{tabular}{@{\hspace{-1em}} c @{\hspace{1em}} c @{\hspace{-1em}} c}
		&\pgfimage[width=.55\textwidth]{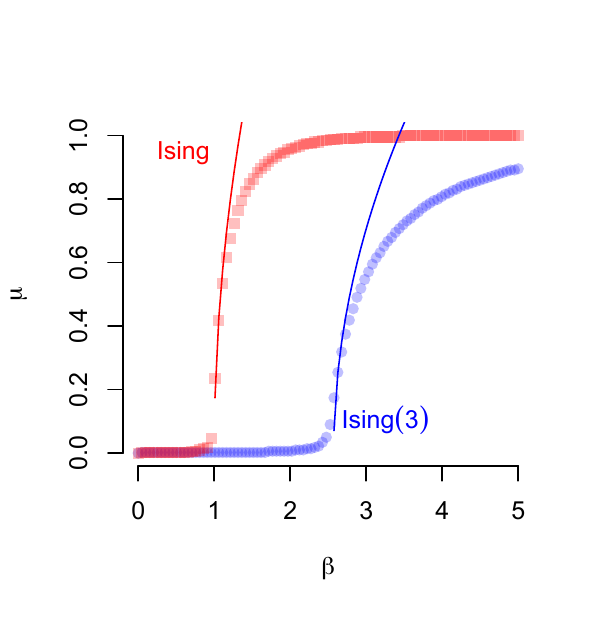}		&\pgfimage[width=.55\textwidth]{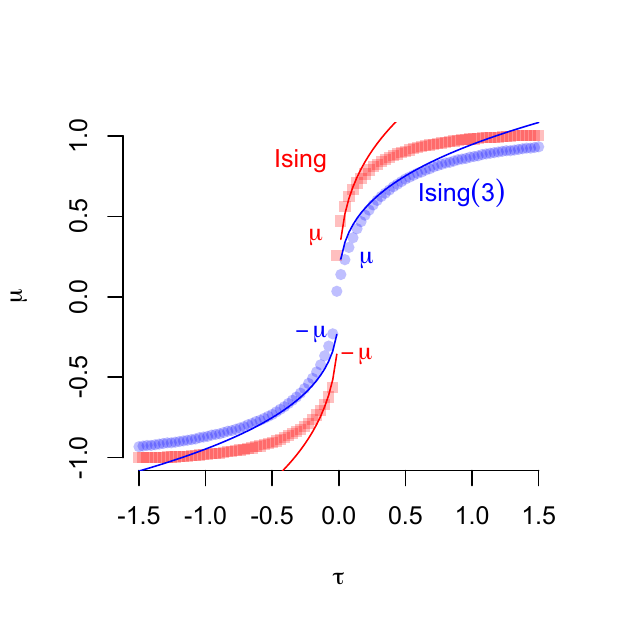}\\[-1em]
	&	(a)		&(b)
\end{tabular}
\caption{ Magnetisation $\mu$ using the mean field with parameters  $d=10$,  $\sigma=0.1$, $\tau=0$ and $k=3$ for the general-spin Ising model (blue circles) and the regular Ising model (red squares). (a) is the magnetisation as a function of the inverse temperature $\beta$. In (b) is the magnetisation as a function of the external field $\tau$. The continuous lines are the predictions from equations (\ref{eq:phase-transition-temp}) in (a) and (\ref{eq:phase-transition-external}) in (b). }
\label{fig:magnetisation}
\end{figure}

In Figure \ref{fig:magnetisation}(a) we show the magnetisation as a function of the inverse temperature, for both the general-spin Ising model with $k=3$ and the regular Ising model. While spontaneous magnetisation emerges in a similar way in both models,  the critical value of the inverse temperature for the general-spin Ising model  is shifted to the right by a factor of $r=(2k+1)/6k$. 

The phase transition of the magnetisation $\mu$ as a function of the external field $\tau$ is obtained 
by considering its behaviour near the critical inverse temperature, which for the general-spin Ising model is $\beta_c=1/(r \sigma d)$. We obtain the following equation (Appendix \ref{app:landau-theory}): 
\begin{align}\label{eq:phase-transition-external}
    \mu\approx\text{sign}(\tau)\left(\frac{|\tau|}{r\sigma d}\right)^{1/3}.
\end{align}
This result can be leveraged to yield the decreasing width of the hysteresis. We expect a transition from $-1$ to $+1$ (or vice versa) to happen at $\tau\approx \pm r\sigma d$ (and equals $\pm1/\beta_c$). And since $r$ will be approximately $\tfrac{1}{3}$ for large values of $k$, we expect that the width of the hysteresis converges. 

Figure \ref{fig:magnetisation}(b) shows the phase transition for the general-spin Ising model and the regular Ising model as a function of the external field $\tau$. As before, the points represent spontaneous magnetisation, where there are two attractive fixed points at the extremes determined by the external field $\tau$. The lines represent the function of equation (\ref{eq:phase-transition-external}); without the factor $r$ for the regular Ising model. We find that  the mean field for the general-spin Ising model with $k=3$ undergoes a similar first-order phase transition but the magnetisation as a function of $\tau$, but does not exhibit a jump from one extreme to the other as large as  what happens to the regular Ising model. The fact that the general-spin Ising model has multiple spin values makes this possible. Again, the factor $r=(2k+1)/6k$ plays a key role in determining spontaneous magnetisation.  

\begin{figure}\centering
\begin{tabular}{@{\hspace{-1em}} c @{\hspace{1em}} c @{\hspace{-1em}} c}
		&\pgfimage[width=.55\textwidth]{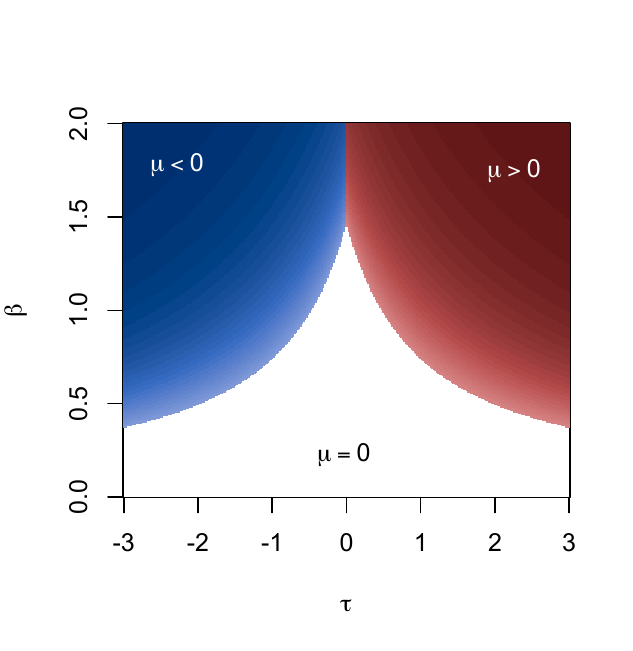}		&\pgfimage[width=.55\textwidth]{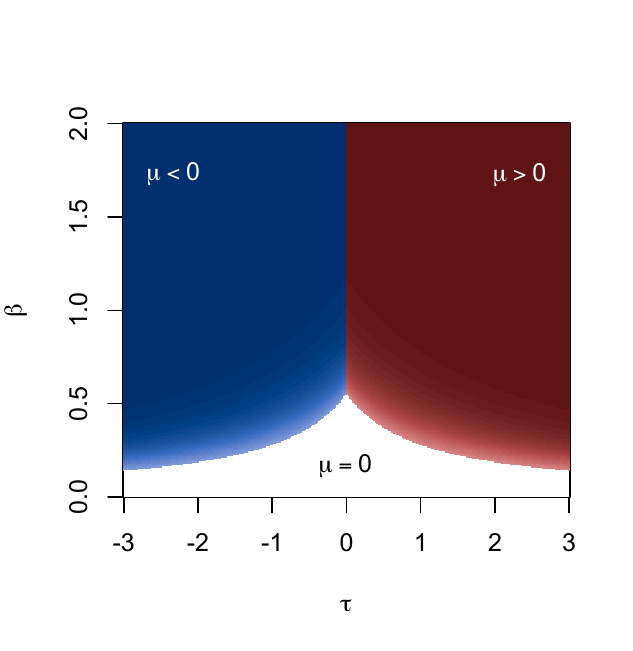}\\[-1em]
	&	(a)		&(b)
\end{tabular}
\caption{Phase maps for the Ising$(3)$ model in (a) and the Ising model in (b). Shown in colors is the spontaneous magnetisation of the mean field as a function of the external field $\tau$ and inverse temperature $\beta$ from equation (\ref{eq:mean-field}) with parameter settings  $d=2$, $\sigma=1$, and $k=3$ in (a). Red color indicates positive magnetisation, blue negative magnetisation and white $0$ magnetisation.   }
\label{fig:phase-maps}
\end{figure}

In Figure \ref{fig:phase-maps}(a) we show the phase map in the $\beta-\tau$-plane for the general-spin Ising model, and in (b) -- that for the regular Ising model. For low values of the inverse temperature $\beta$, the spins are approximately random and so the magnetisation is $\mu=0$. For larger values of $\beta$ spontaneous magnetisation arises. Edges of the phase maps show the combination of $\beta$ and $\tau$ where spontaneous magnetisation arises, and remains occurring in the coloured areas; blue for negative magnetisation and red for positive magnetisation. Comparing the general-spin Ising model and the regular Ising model, the general-spin Ising model requires lower temperatures for spontaneous magnetisation than the regular Ising model. 

Summarising, the mean-field behaviour of the general-spin Ising model shows very similar behaviour to the regular Ising model, but depends on an additional factor $r=(2k+1)/6k$, with $k\in \mathbb{N}$.

\section{Numerical illustration}\label{sec:numerical-illustration}\noindent
To determine that the predictions from the mean field solution in (\ref{eq:mean-field}) and its consequences for magnetisation in (\ref{eq:phase-transition-temp}) and (\ref{eq:phase-transition-external}) are accurate, we perform Monte Carlo simulations. We use the single-site Metropolis algorithm \citep[][see Appendix \ref{app:metropolis-algorithm}]{Metropolis:1953}. 
An Erd\"{o}s-R\'{e}nyi network is generated with varying probability of an edge $p_e$ and varying number of nodes $n$, so that the average degree $d$ also varies. Connectivity is $\sigma=0.1$. 

\begin{figure}\centering
\begin{tabular}{@{\hspace{-1em}} c @{\hspace{-1.5em}} c @{\hspace{-1.5em}} c}
		\pgfimage[width=.40\textwidth]{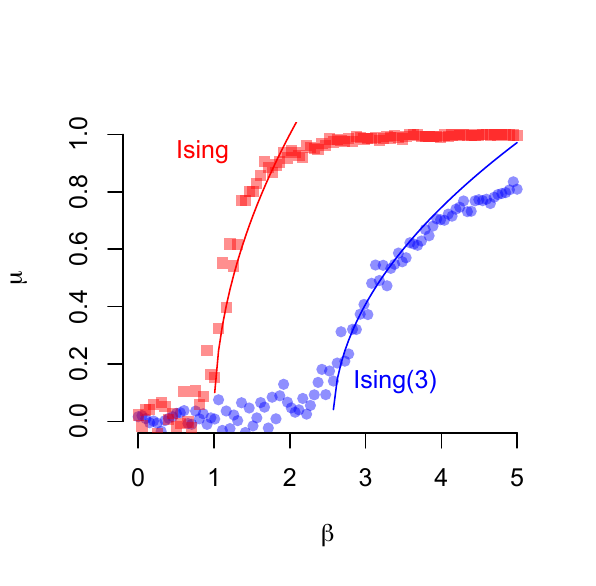}		&\pgfimage[width=.38\textwidth]{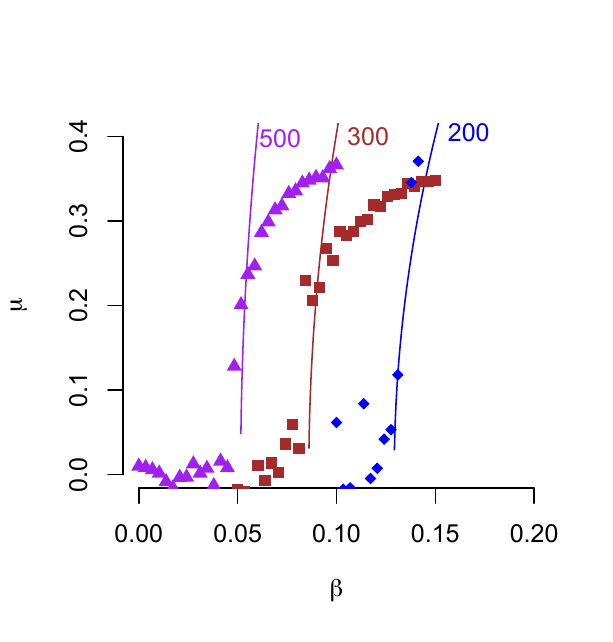}		&\pgfimage[width=.38\textwidth]{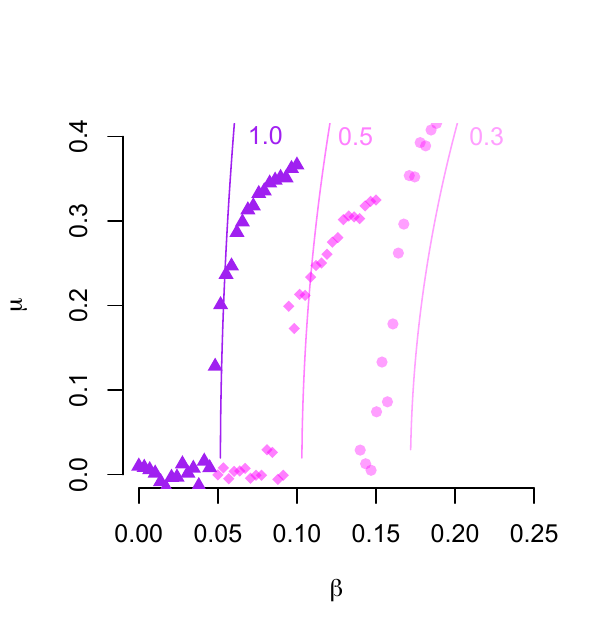}\\[-1em]
	(a)		&(b)  &(c)
\end{tabular}
\caption{  In (a) the magnetisation $\mu$ as a function of $\beta$ of the regular (red squares) and the general-spin Ising model with $k=3$ (blue circles) obtained with the Metropolis sampler and the theoretical values obtained with (\ref{eq:phase-transition-temp}) as solid lines. The parameters are $n=100$, $\sigma=0.1$, $p_e=0.1$.  In (b) is the magnetisation (\ref{eq:phase-transition-temp}) a function of $\beta$ for different size of the complete graph with $n=200$, $300$, and $500$ nodes.  In (c) is the magentisation (\ref{eq:phase-transition-temp}) as a function of $\beta$ for a random graph with $n=500$ nodes and edge probability $p_e=1$, $0.5$, and $0.3$.     }
\label{fig:num-magnetisation-ising}
\end{figure}

In Figure \ref{fig:num-magnetisation-ising}(a) is the magnetisation $\mu$ as a function of $\beta$. The location of the spontaneous magnetisation of the general-spin Ising model (blue circles) confirms the theoretical predictions (solid line).
Near criticality, we see in Figure \ref{fig:num-magnetisation-ising}(b) that the accuracy of the mean field becomes worse with smaller graphs, and that this dependency also holds when the degree of each node is decreased \citep[Figure \ref{fig:num-magnetisation-ising}(c); we left in $p_e=0.3$ to use for sparse graphs, as used in belief propagation on tree-like graphs, e.g., ][]{Castro:2025}.  

\begin{figure}\centering
\begin{tabular}{@{\hspace{-1em}} c @{\hspace{-1.5em}} c @{\hspace{-1.5em}} c}
    \pgfimage[width=.40\textwidth]{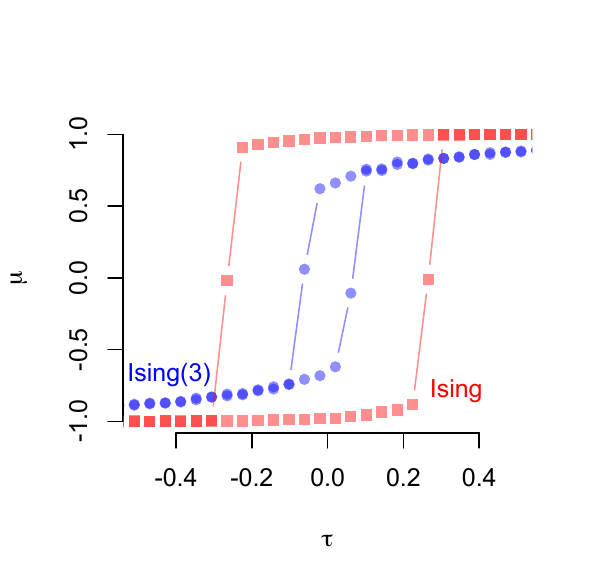}
    &\pgfimage[width=.38\textwidth]{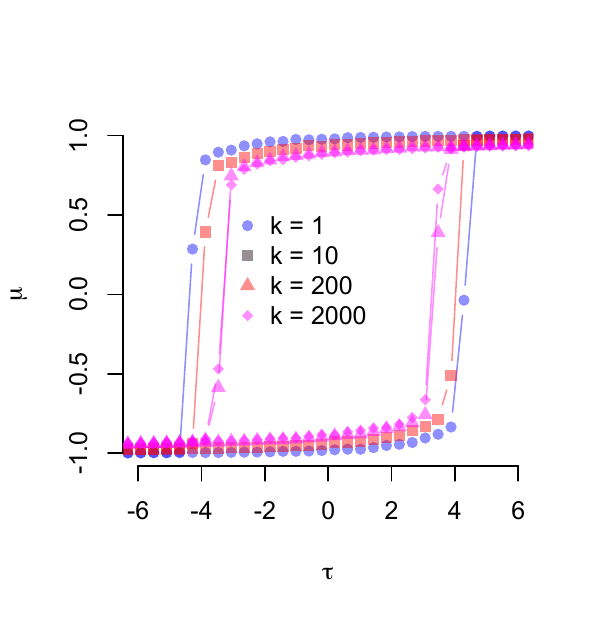}		&\pgfimage[width=.38\textwidth]{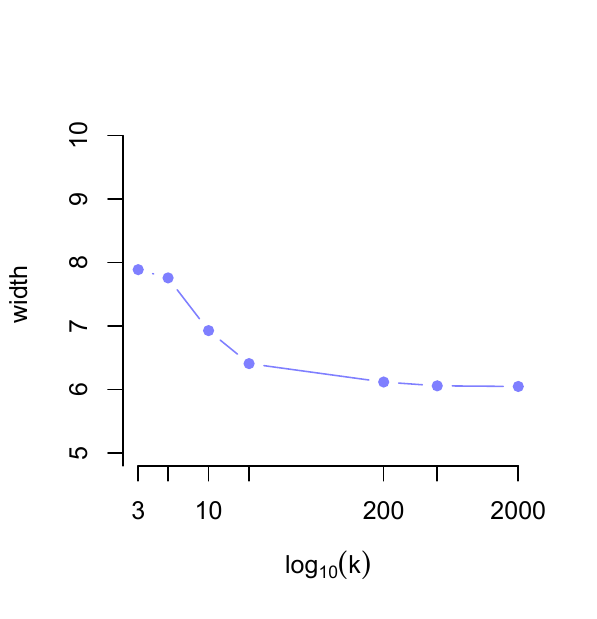}\\[-1em]
	(a)		&(b)  &(c)
\end{tabular}
\caption{  In (a) is the magnetisation as a function of the external field $\tau$ and with $\beta=5$. Increasing $\tau$ from $-0.5$ to $0.5$ (circles), and going from $0.5$ to $-0.5$ (squares). In (b) are hysteresis plots for the general-spin Ising model for different values of $k=1$, $10$ and $200$ and $2000$. In (c) are the widths of the hysteresis for $k=3$ up to $k=2000$ in $\log$ with base $10$. Other parameters are $\beta=1.1$, $n=100$, $p_e=0.1$.  }
\label{fig:num-magnetisation-categ}
\end{figure}

Figure \ref{fig:num-magnetisation-categ}(a) shows hysteresis of the magnetisation $\mu$ as a function of the external field $\tau$. This effect is because of the memory (residual) magnetisation when changing the external field. Comparing with the regular Ising model, the hysteresis effect is less strong (the upward and downward jumps are closer) for the general-spin Ising model than for the regular Ising model.

Figure \ref{fig:num-magnetisation-categ}(b) shows that there is a strong indication that the hysteresis effect converges to a particular point. It is easily proved that if each term $\sinh(x\beta\gamma)$ in the numerator and $\cosh(x\beta\gamma)$ in the denominator of the mean field expression in (\ref{eq:mean-field}) is $<1$ (in absolute value), then with $k\to \infty$ the mean field will converge. This is sufficient because the partial series $\sum_{s=1}^k \tfrac{s}{k}$ has ratio test value 1. Figure \ref{fig:num-magnetisation-categ}(c) confirms this numerically. 

\section{Conclusion and discussion}\label{sec:conclusion-discussion}\noindent
The general-spin Ising model, a model with multiple valued spins, symmetric around zero between $-1$ and $+1$, is important not only as a theoretical framework, but also because of its applications—ranging from $S$-spin models in physics to polarization dynamics in sociology and attitude modeling in psychology. In such applications, it is often valuable to understand the types of states and dynamics the model can exhibit, in order to assess its suitability for a given modeling context. Mean-field theory offers insights into these properties.

 We obtained the Gibbs free energy for the mean field and minimised this to obtain the mean field solution. The properties of the general-spin mean field are similar to those of the regular Ising model, i.e., the emergence of spontaneous magnetisation via a second-order transition at low temperatures, and hysteresis. However, spontaneous magnetisation occurs at a shifted value compared to the regular Ising model, determined by $r=(2k+1)/6k$, where $2k+1$ denotes the number of categories of the general-spin Ising model. The phase diagram showed that temperatures need to be lower to obtain spontaneous magnetisation in the general-spin Ising model than for the regular Ising model. Additionally, the hysteresis effect is slightly different depending on the number of categories, but converges when taking the limit of  $k\rightarrow \infty$.  Monte Carlo simulations confirm the theoretical predictions derived from the mean field, and indicate the dependence on the system size and node degree. 

Using these results allows for the application of the general-spin Ising model in different fields. In psychology, for example, this model is applicable in contexts where responses are measured using a visual analogue scale, typically implemented as a slider with around 100 response categories. These types of responses could then be modelled with a general-spin Ising model with many categories. Experiments should reveal whether the predictions from mean-field theory accurately reflect empirical data.

\section*{Acknowledgement}\noindent
 The research conducted by Han L. J. van der Maas has been supported by a grant from the ERC (ERC project 101053880 – CASCADE). The research conducted by Tuan Pham is supported by the Dutch Institute for Emergent Phenomena (DIEP) cluster at the
University of Amsterdam under the Research Priority Area \emph{Emergent Phenomena in Society: Polarisation, Segregation and Inequality}. The research by L. Waldorp has been supported by the "New Science of Mental Disorders", the Dutch Research Council and the Dutch Ministry of Education, Culture and Science
(NWO), grant number 024.004.016.

\section*{Data Availability Statement}\noindent
This manuscript does not contain data. 

\section*{Author contributions}\noindent
LW, TP and HvdM developed the ideas and concepts, derived results and wrote the manuscript. LW performed the simulations.

\appendix
\section*{Appendix}
\section{Mean field derivation}\label{app:mean-field}\noindent
Intuitively, the mean field assumes that in a well enough connected graph $\mathcal{G}$,  the links  are weak \citep[see][]{Gleeson:2012}, such that  we can obtain the  measure  $\mathbb{P}_\theta(x)$ in Eq. \eqref{eq:ising-model} as a product of the marginal distributions $\mathbb{P}^{(i)}_\theta(x_i)$, and so that the partition function is a sum of products \citep{Plischke:1994,Gleeson:2012}.
To construct the mean field we need two assumptions. These assumptions are similar to those described by, e.g., \citet{Plischke:1994}. 

\begin{assumption}{(Large degree)}\label{ass:uniform-graph}
Each node in the graph $\mathcal{G}$ has sufficiently high degree $d$.
\end{assumption}
In practice we use a random (Erd\"{o}s-Reny\'{i}) graph with sufficiently high probability of connection $p_e$, so that approximately $d=p_e(n-1)$ for each node. 

\begin{assumption}{(Concentration)}\label{ass:concentration}  The nodes (variables) $X_s$ in graph $\mathcal{G}$ are sub-gaussian random variables with mean $\mu$ and variance proxy $\sigma^2$.  
\end{assumption}
Assumption \ref{ass:concentration} is another (quantitative) way to ensure that the correlations between nodes remains low (see below); it implies that for any $s\in V$ we obtain by Hoeffding's bound \citep{Vershynin:2018}
\begin{align*}
\mathbb{P}(|X_s-\mu|\ge \delta n)\le 2\exp\left( -\frac{\delta^2 n^2}{\sigma^2} \right),
\end{align*}
where $n$ is the number of nodes in $\mathcal{G}$, i.e., $n=|V|$. This concentration of measure is required for the mean field approximation, to keep the correlations between the variables small. Because the random variables $X_s$ are bounded between $-1$ and $+1$, we immediately obtain that they are sub-gaussian with variance proxy 1 \citep{Vershynin:2018}.

To obtain the mean field we obtain the free energy and use the first-order derivative to obtain the first moment \citep[e.g.,][]{McCulloch:1988}. We have the partition function 
\begin{align*}
Z=\sum_{x\in \Omega_k}\exp\left(\beta\tau\sum_{s\in V} + \beta\sigma\sum_{(s,t)\in E} x_s x_t\right).
\end{align*}
The interactions for random variables $X_s$ can be rewritten as
\begin{align*}
    X_s X_t=(X_s-\mu+\mu)(X_t-\mu+\mu)=(X_s-\mu)(X_t-\mu)+2\mu(X_s-\mu)+\mu^2.
\end{align*}
By Assumption (\ref{ass:concentration}) the covariance term is small. Assuming that the random variables $X_s$ are sub-gaussian implies that 
\begin{align*}
    \mathbb{P}(|(X_s-\mu)(X_t-\mu)|\ge \delta n)\le 2 \exp\left(-\frac{n^2\delta^2}{2\sigma^2}+\log 2\right).
\end{align*}
Therefore, taking the right hand side as $\epsilon$, so that $\delta=\sqrt{\tfrac{2\sigma^2}{n^2}\log \tfrac{1}{\epsilon}}$, we obtain with high probability $1-\epsilon$ that
\begin{align*}
    |(X_s-\mu)(X_t-\mu)|\le  \sqrt{2\sigma^2\log\frac{1}{\epsilon}},
\end{align*}
and we see that the covariance terms are small. Relatedly, this result can be connected to the Ginzburg criterion \citep[see e.g.,][]{Plischke:1994}, such that we demand that $\epsilon\ge \exp( -\frac{\mu^4}{2\sigma^2})$, i.e., we demand that the mean field is large enough in the fourth order moment compared to the second order moment, i.e., the second-order fluctuations cannot be too large.

And so we have the mean field Hamiltonian
\begin{align*}
\mathcal{H}_{\mu}(x)=-\tau\sum_{s\in V} x_s-\sigma\sum_{(s,t)\in E}(-\mu^2 +2\mu x_s).
\end{align*}
Noting that $\sum_{(s,t)}\mu^2=\mu^2 n d/2$, with $d$ the degree of each node in $\mathcal{G}$ (by Assumption \ref{ass:uniform-graph}), and $\sum_{(s,t)} x_s= (d/2) \sum_s x_s$. We then obtain the mean field Hamiltonian
\begin{align*}
\mathcal{H}_{\mu}(x)=\sigma\mu^2 n \frac{d}{2} -(\sigma\mu d +\tau)\sum_{s\in V} x_s.
\end{align*}
The partition function for the mean field $Z_\mu$ then becomes
\begin{align*}
Z_\mu
&=
\exp(-\beta\sigma\mu^2 n d/2) \sum_{x_s\in \Omega_k}\exp\left(\beta(\sigma \mu d+\tau)\sum_{s\in V} x_s\right),
\end{align*}
By Assumption \ref{ass:uniform-graph} we see that we can replace each $x_s$ by a generic $x$. Hence, we obtain
\begin{align*}
Z_\mu
&=
\exp(-\beta\sigma\mu^2 n d/2)\left(\sum_{x\in \Omega_k^+}2\cosh\left(\beta(\sigma\mu d+\tau)x\right)\right)^n,
\end{align*}
where $\Omega_k^+=\{0,\tfrac{1}{k},\tfrac{2}{k},\ldots,1\}$. We can then define the Gibbs free energy for the mean field as 
\begin{align*}
G_\mu=-\frac{1}{\beta}\log Z_\mu =
\sigma\mu^2 n \frac{d}{2} - \frac{n}{\beta}\log\sum_{x\in \Omega_k^+}2\cosh\left(\beta(\sigma\mu d+\tau)x\right),
\end{align*}
To obtain the mean field we need to minimise the free energy $G_\mu$, leading to 
\begin{align*}
    \mathbb{E}(x) 
    &=\mu 
    =
    \frac{\sum_{x\in \Omega_k^+}x\sinh(x\beta\gamma)}
            {\sum_{x\in \Omega_k^+}\cosh(x\beta\gamma)}
\end{align*}
where $\gamma=\sigma\mu d +\tau$.

\section{Landau theory}\label{app:landau-theory}\noindent
We consider Landau theory for the mean field in (\ref{eq:mean-field}). Initially, we set the external field $\tau=0$. The approximation is obtained by a Taylor series of (\ref{eq:mean-field}) at $\mu=0$
\begin{align*}
    \mu=a_0 + a_1\mu + a_2\mu^2+a_3\mu^3 +O(\mu^4).
\end{align*}
The first coefficient is $a_0$, which can be seen to be 0 by setting in (\ref{eq:mean-field}) $\mu=0$ in $\gamma$ and recalling that $\tau=0$. For $a_1$ we require the first derivative. Let $C=1+\cosh(\tfrac{1}{k}\beta\gamma)+\cosh(\tfrac{2}{k}\beta\gamma)+\cdots +\cosh(\beta\gamma)$. Then the derivative of (\ref{eq:mean-field}) gives $a_1$
\begin{align*} 
        \beta\sigma d\frac{C\sum_{x\in \Omega_k^+}x^2\cosh(x\beta\gamma)}
            {C^2}  
        =
        \beta\sigma d\frac{(\frac{1}{k})^2+(\tfrac{2}{k})^2+\cdots +1}{k+1}
        =
        \beta\sigma d\frac{2k+1}{6k},
\end{align*}
where we already set $\mu=0$ so that all $\sinh$ terms are 0, and we used the fact that $\sum_{x\in \Omega_k^+}x^2=(k+1)(2k+1)/6k$.
It is obvious that $a_2=0$ since all terms contain $\sinh$. Then, the next derivative gives 
\begin{align*}
a_3= (\beta\sigma d)^3\frac{C\sum_{x\in \Omega_k^+}x^4\cosh(x\beta\gamma)} {C^2} 
-3(\beta\sigma d)^3\frac{\left(\sum_{x\in \Omega_k^+}x^2\cosh(x\beta\gamma)\right)^2}{C^2},
\end{align*}
where we again already removed the $\sinh$ terms which are 0 at $\mu=0$. Then we obtain 
\begin{align*}
a_3= (\beta\sigma d)^3\frac{\sum_{x\in Q_+}x^4\cosh(x\beta\gamma)} {C^2} 
-3(\beta\sigma d)^3\frac{\left(\sum_{x\in Q_+}x^2\cosh(x\beta\gamma)\right)^2}{C^2}. 
\end{align*}
Using that $\sum_{x\in [n]} x^4=k(k+1)(2k+1)(3k^2+3k-1)/30k^4$, we get
\begin{align}\label{eq:a3}
a_3= (\beta\sigma d)^3\left(\frac{k(2k+1)(3k^2+3k-1)}{30(k+1)^3}-3\frac{r^2}{(k+1)^2}\right),
\end{align}
where again we have set $\mu$ to 0 so that $\sinh$ terms drop out and where $r=(2k+1)/6k$.  
This gives the approximation
\begin{align*}
    \mu= r\beta \sigma d \mu - \frac{1}{3}a_3\mu^3 + O(\mu^5).
\end{align*}
And so, approximately
\begin{align*}
    \mu\approx\pm a_3^{-1/2}\sqrt{3}\sqrt{r\beta \sigma d-1}.
\end{align*}
So, there is a second-order transition when $r\beta\sigma d$ approaches 1 from above, and if $r\beta \sigma d\le  1$, then the magnetisation is 0.

We can obtain a similar result when including the  external field. In the above Taylor expansion we  use $r\beta_c\sigma\mu d +\beta_c\tau$, where $\beta_c$ is the critical inverse temperature $(r\sigma d)^{-1}$, so that we have $\mu +\beta_c\tau$ \citep[][Section 3.6]{Plischke:1994}. Then we obtain 
\begin{align*}
    \mu= \mu + \beta_c\tau - \frac{1}{3}a_3(\mu +\beta_c\tau)^3 + O(\mu^5).
\end{align*}
And we obtain that the magnetisation changes according to the external field approximately as
\begin{align*}
    \mu\approx\text{sign}(\tau)\left(\frac{3|\tau|}{a_3 r\sigma d}\right)^{1/3}.
\end{align*}
%

\newpage
\section{Metropolis algorithm}\label{app:metropolis-algorithm}
\begin{algorithm}[h]
\caption{Metropolis algorithm}
Recall that $V=\{1,2,\ldots,n\}$ is the set of nodes and  $Q=\{-1,-\tfrac{k+1}{k},\ldots,0,\tfrac{1}{k},\tfrac{2}{k},\ldots,1\}$ is the set of states of $X$.
\begin{algorithmic}
\State $i \gets 1$
\For{$i\le niter$} 
  \For{$v$ in the set of nodes $V$}
    \State sample $v\in V$ with equal probability $\frac{1}{|V|}$
    \State propose state $x_v'$ as one of $Q\backslash \{x_v\}$ with uniform probability
    \State leave all other nodes $j\in V\backslash \{v\}$ as is
    \State obtain Hamiltonians\\ 
    \State $\mathcal{H}(x)=-x_v'\tau_s -x_v'\sum_{j\ne s} \sigma_{sj} x_j$ and \\
    \State $\mathcal{H}(x')=-x_v'\tau_s -x_v'\sum_{j\ne v} \sigma_{sj} x_j$\\
    \State compute difference $\mathcal{H}(x')-\mathcal{H}(x)$
    \State obtain $u$ uniformly from $[0,1]$
    \State set $p=\exp(-\beta(\mathcal{H}(x')-\mathcal{H}(x)))$
    \If{$\min\{1,p\} > u$} $x_v \leftarrow x_v'$
    \ElsIf{$\min\{1,p\}\le u$} $x_v$ remains the same
     \EndIf
  \EndFor
  \State $i \leftarrow i+1$
\EndFor
\end{algorithmic}
\end{algorithm}
%


\end{document}